\documentclass{svproc}
\usepackage{wrapfig}
\usepackage{float}
\usepackage{url}
\usepackage{hyperref}
\usepackage{xspace}
\usepackage{color}
\usepackage{xcolor}
\usepackage{rotating}
\definecolor{light-gray}{gray}{0.8}

\newcommand{\pPb}{p--Pb\xspace}
\newcommand{\PbPb}{Pb--Pb\xspace}

\newcommand{\GeVc}{\ensuremath{{\rm GeV/}c}\xspace}

\newcommand{\sqrts}{\ensuremath{\sqrt{s}}\xspace}

\newcommand{\sqrtsNN}{\ensuremath{\sqrt{s_{\rm NN}}}\xspace}
\newcommand{\pt}{\ensuremath{{\it p}_{\rm T}}\xspace}

\newcommand{\cmnt}[1]{}

\usepackage{lineno}
\usepackage{amssymb}
\usepackage{amsmath}
\usepackage{bm}
\begin{document}
\mainmatter              % start of a contribution
\title{Heavy-flavour jet production and charm fragmentation with ALICE\thanks{A Large Ion Collider Experiment} at LHC\thanks{Large Hadron Collider}}
\titlerunning{Heavy-flavour jets with ALICE}  % abbreviated title (for running head)
\author{Auro Mohanty\inst{*} on behalf of the ALICE Collaboration}%

\institute{Institute for Subatomic Physics, Utrecht University/Nikhef, Utrecht, Netherlands\\
\email{auro.mohanty@cern.ch}}

\maketitle              % typeset the title of the contribution
\begin{abstract}%\vspace{-0.6cm}
Heavy quarks, produced in hard parton scatterings in the early stage of ultra-relativistic heavy-ion collisions, are ideal probes to investigate the properties of the Quark--Gluon Plasma (QGP) produced in such collisions. Measurements of heavy-flavour jets can provide constraints on energy-loss models. In particular, they add information on how the radiated energy is dissipated in the medium. 
Studies of angular correlations between heavy-flavour and charged particles allow us to characterize the heavy-quark fragmentation process and its possible modification in a hot nuclear matter environment. 

This manuscript will focus on the latest results on heavy-flavour jets and D-meson correlations with charged particles studied with the ALICE detector in pp, p--Pb and Pb--Pb collisions.
 
\keywords{Heavy-flavour, jets, nuclear modification,  fragmentation function, momentum fraction}
\end{abstract}
\section{Physics Motivation}\label{intro}
Fragmentation function  
of photon-triggered mesons was studied by Kang and Vitev~\cite{Kang:2011rt} 
and a flavour dependence of energy loss in QGP medium was predicted. Measurements in pp collisions provide essential reference to interpret those in proton--nucleus (p--A) and nucleus--nucleus (A--A) collisions. They also provide an excellent test of the perturbative quantum chromodynamics (pQCD) because heavy-flavour observables are calculable in pQCD down to $p_{\text T}$ $\approx$ 0. Anderle et al.~\cite{Anderle:2017cgl} presented a global QCD analysis of D$^{*\pm}$-meson fragmentation functions in pp scatterings. 

ALICE is uniquely placed to play a significant role in the low and intermediate $p_{\text T}$ (\pt: transverse momentum) sector. 
%%%%%%%%%%%%%%%%%%%%%%%%%%%%%%%%%%%
\section{Procedure and Physics Results}
Heavy-flavour jets are studied by means of two different methods: by reconstructing jets with a heavy-flavour tag (`heavy-flavour jets') and by studying correlations between heavy-flavour hadrons with other hadrons (`D-meson-hadron correlations').

\subsection{Heavy-flavour jets}
	Jets are reconstructed using the anti-$k_{\text T}$ algorithm~\cite{Cacciari:2008gp}. They are tagged as heavy-flavour jets if they have within their constituents: heavy-flavour electrons, D mesons, or beauty mesons (by an indirect measurement).

%%%%%%%%%%%%%%%%%%%%%%%%
\vspace{-0.1cm}
	\subsubsection{Heavy-flavour electron jets.}	
	Electrons resulting from the semi-leptonic decay of heavy-flavour hadrons are used to tag the jets, called heavy-flavour electron (HFe) jets. First, jets are reconstructed using charged tracks. Then a constituent track in each jet is searched for, having the same momentum as the heavy-flavour electrons identified separately (see Ref.~\cite{Collaboration_2008} for a detailed description of the ALICE apparatus).
	
         \begin{figure}[!h]%{0.4\textwidth}\vspace{-0.5cm}
	\begin{center}\vspace{-0.5cm}
	\includegraphics[width=0.5\linewidth]{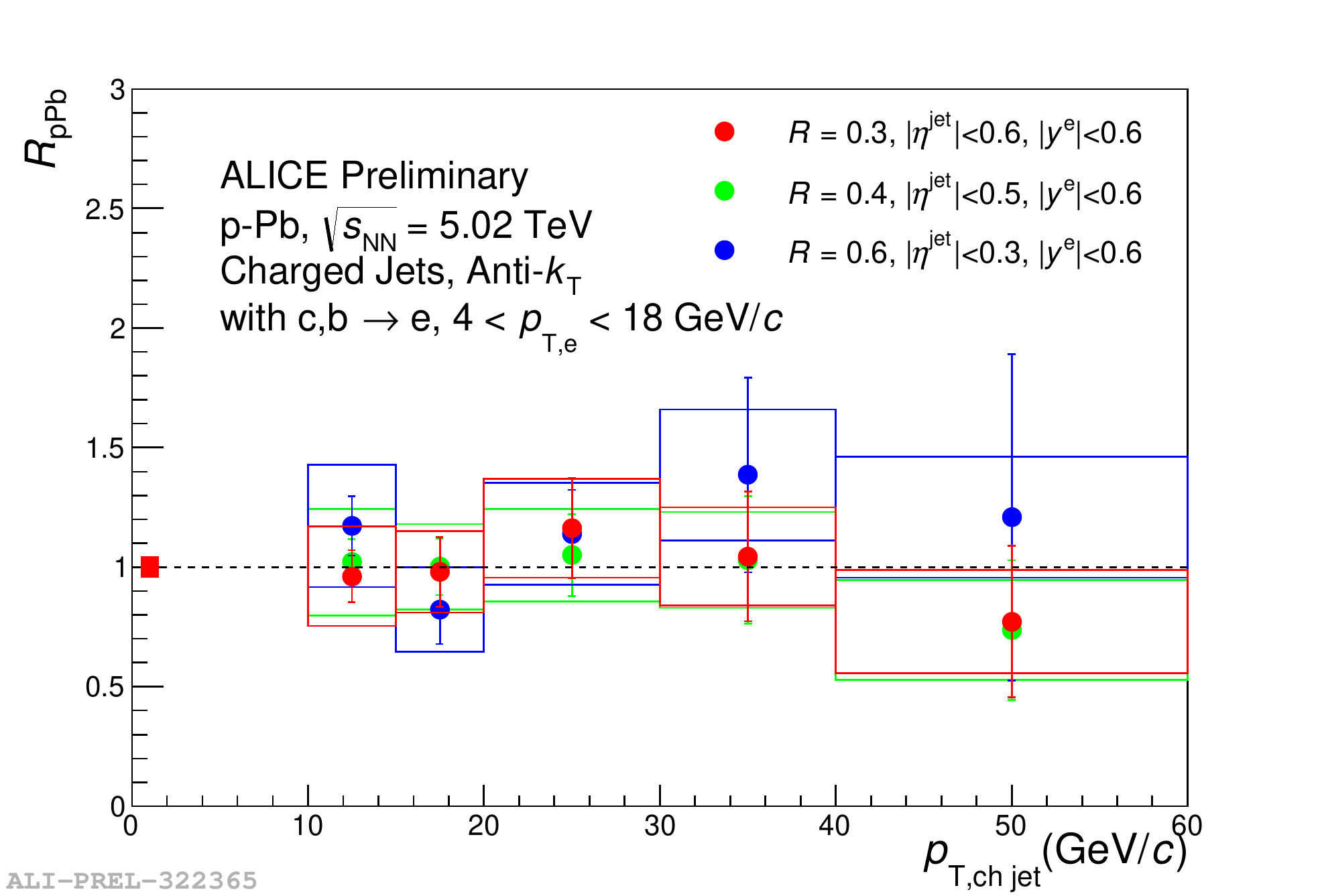}
	\end{center}\vspace{-0.65cm}
	\caption{$R_{\text{pPb}}$ of HFe jets with $R=0.3, 0.4,$ and $0.6$.}
	\label{fig:HFejets-shower}
         \end{figure}%\vspace{-0.4cm}
         
         Figure \ref{fig:HFejets-shower} shows the nuclear modification factor $R_{\text{pPb}}$ of HFe jets with jet radii $R=0.3, 0.4,$ and $0.6$ in p--Pb collisions at a center of mass energy per nucleon pair, $\sqrt{s_\text{NN}}=5.02$ TeV. No cold nuclear matter (CNM) effects are observed.
%%%%%%%%%%%%%%%%%%%%%%%%
\vspace{-0.1cm}
	\subsubsection{D-meson tagged jets.}
	Jets are tagged if they contain a D$^0$ meson within the jet cone. D$^0$ mesons are reconstructed in the D$^0 \rightarrow$ K$^-$ $\pi^+$ ~\cite{PhysRevD.98.030001} decay channel. The daughter kaon and pion tracks are replaced by an equivalent D$^0$ constituent, which is then used together with the other charged tracks to reconstruct the jets. 
	
\begin{figure}[!h]%[13]{l}{0.7\textwidth}
\hspace{0.2cmm}	\begin{minipage}{0.39\hsize} 
	\begin{center}
	\includegraphics[width=1\linewidth]{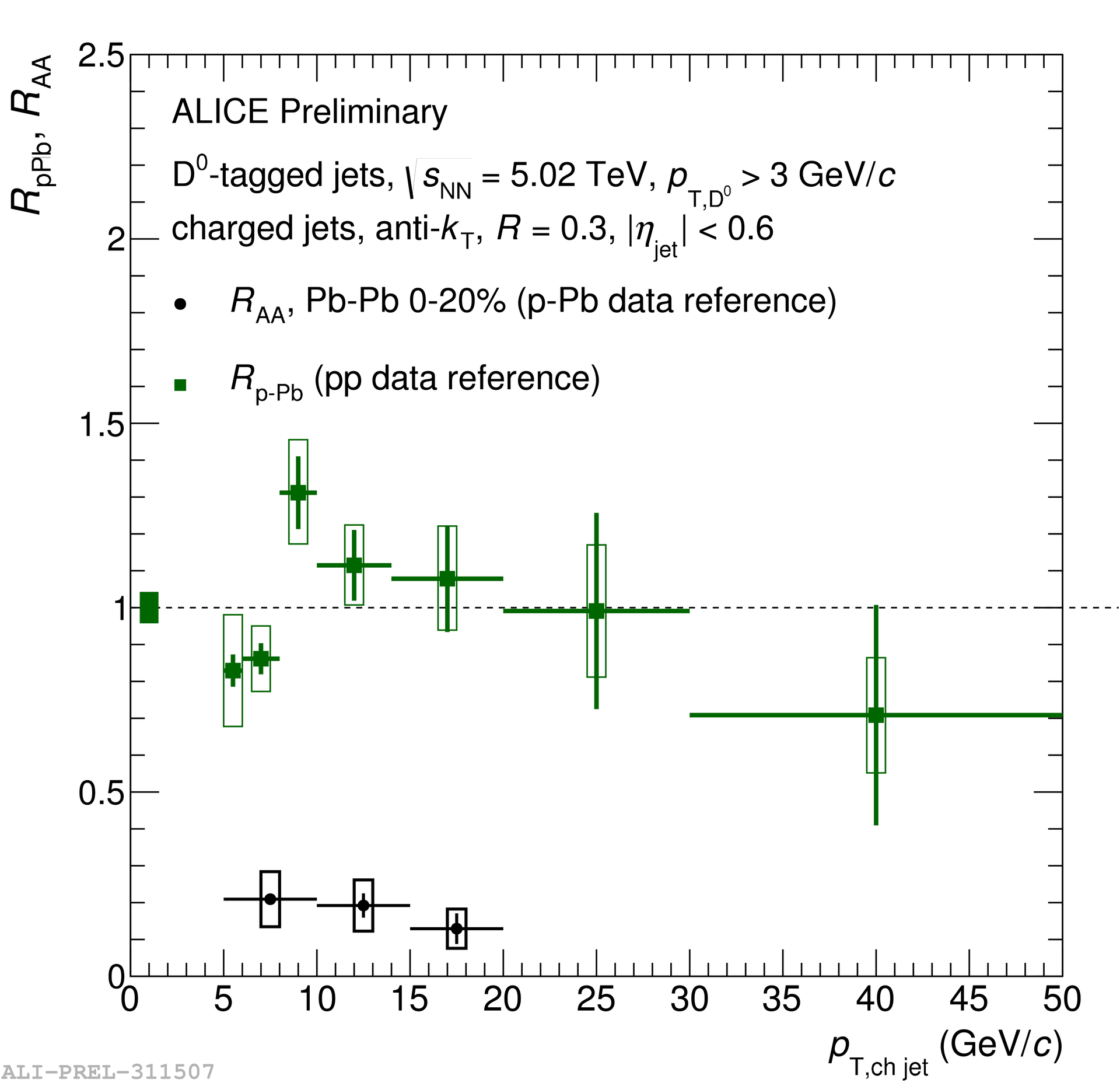}
	\end{center}
	\end{minipage}\hspace{1.2cm}
	\begin{minipage}{0.39\hsize} 
	\begin{center}
	\includegraphics[width=1\linewidth]{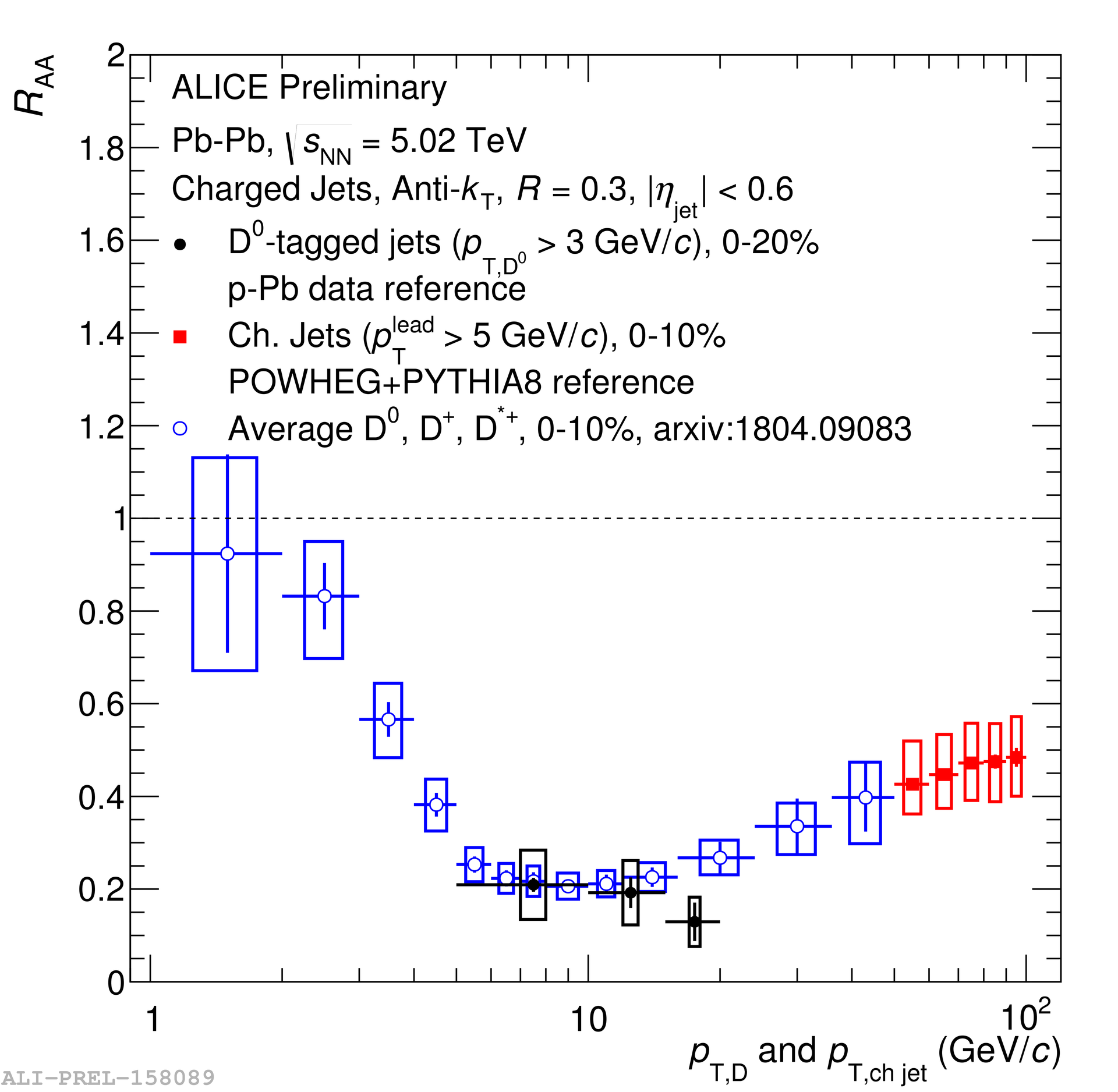}
	\end{center}
	\end{minipage}\vspace{-0.2cm}
	\caption{$R_{\text{pPb}}$ and $R_{\text{AA}}$ of D-jets (left) and $R_{\text{AA}}$ of D-jets, D mesons, and charged jets (right) at \sqrtsNN = 5.02 TeV.} 
	\label{fig:Djets-jetpt}
	\end{figure}
	
	Jet-\pt\ differential cross section of D$^0$-jets was measured in pp collisions at \sqrts = 13 TeV with $R$ = 0.4. The D$^0$-jet production cross section was also measured at \sqrtsNN = 5.02 TeV in pp, \pPb, and \PbPb\ collisions in 0--20\% centrality with $R$ = 0.3. The nuclear modification factor is shown for \pPb\ and \PbPb\ in Fig.~\ref{fig:Djets-jetpt} (left panel). $R_{\text{pPb}}$ is consistent with unity within uncertainties, while $R_{\text{AA}}$ is 0.2 at \pt\ $\sim$ 10 \GeVc\ . On the right panel, it can be seen that the $R_{\text{AA}}$ of D$^0$-jets is compatible with that of D mesons. 
	
\begin{figure}[!h]
\hspace{0.2cm}	
\begin{minipage}{0.4\textwidth} 
	\begin{center}
	\includegraphics[width=\linewidth]{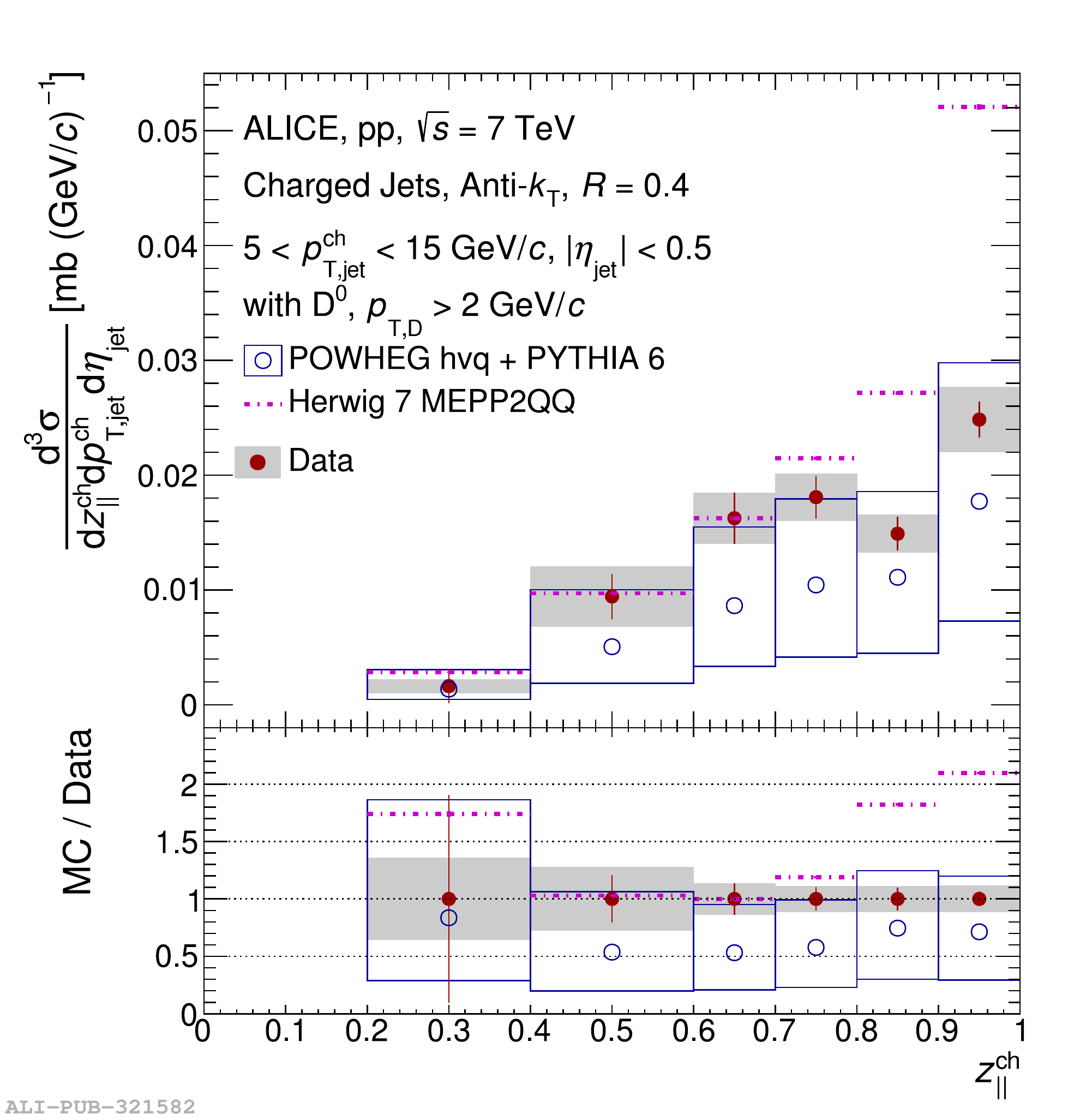}
	\end{center}
	\end{minipage}\hspace{1.2cm}
	\begin{minipage}{0.4\textwidth} 
	\begin{center}
	\includegraphics[width=\linewidth]{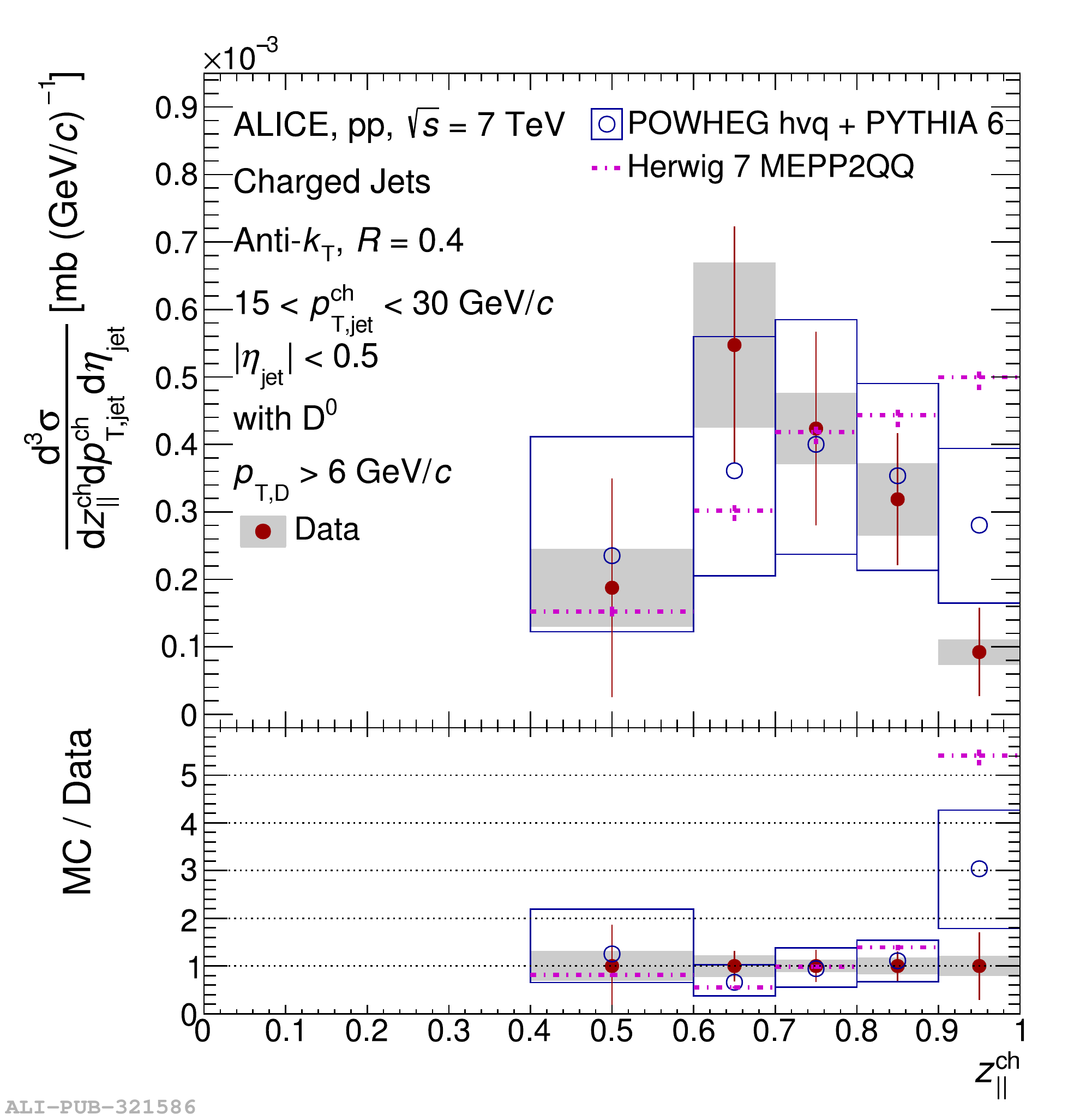}
	\end{center}
	\end{minipage}\vspace{-0.2cm}
	\caption{$z_{||}^{\text{ch}}$-differential cross section of D$^0$-jets in pp collisions at \sqrts = 7 TeV.} 
	\label{fig:Djets-z-7TeV}
	\end{figure}
	
	Fractional momentum ($z_{||}^{\text{ch}}$) carried by the constituent D$^0$ meson along the jet axis was measured for D$^0$-jets~\cite{Acharya:2019zup} in pp collisions at $\sqrt{s}=7$ TeV, with $R$ = 0.4. Hard fragmentation is observed in $5 < $ $p_{\text T}^{\text{ch. jet}}$ $<15$ GeV/$c$, compatible with leading order (LO) and next-to-leading order (NLO) pQCD predictions, as seen on the left in Fig.~\ref{fig:Djets-z-7TeV}. However, in $15 < p_{\text T}^{\text{ch. jet}}$ $< 30$ GeV/$c$, there is a hint of softer fragmentation observed in data, seen on the right panel.
%%%%%%%%%%%%%%%%%%%%%%%%
\subsubsection{b-tagged jets.}
	The property of B mesons having a longer lifetime is exploited here to identify jets originating from b quarks without explicitly reconstructing the B mesons. Jets containing a 3-pronged secondary vertex within the cone are selected since B mesons tend to decay into at least three daughters. Our measurements of jet-\pt\ differential production cross section of b-jets in p--Pb collisions at $\sqrt{s_{\text{NN}}}$=5.02 TeV are in agreement with POWHEG+PYTHIA predictions.
%%%%%%%%%%%%%%%%%%%%%
\vspace{-0.1cm}
	\subsection{D-meson-hadron correlations}
\begin{figure}
\hspace{0.7cm}	
\begin{minipage}{0.38\hsize} 
	\begin{center}
	\includegraphics[width=1\linewidth]{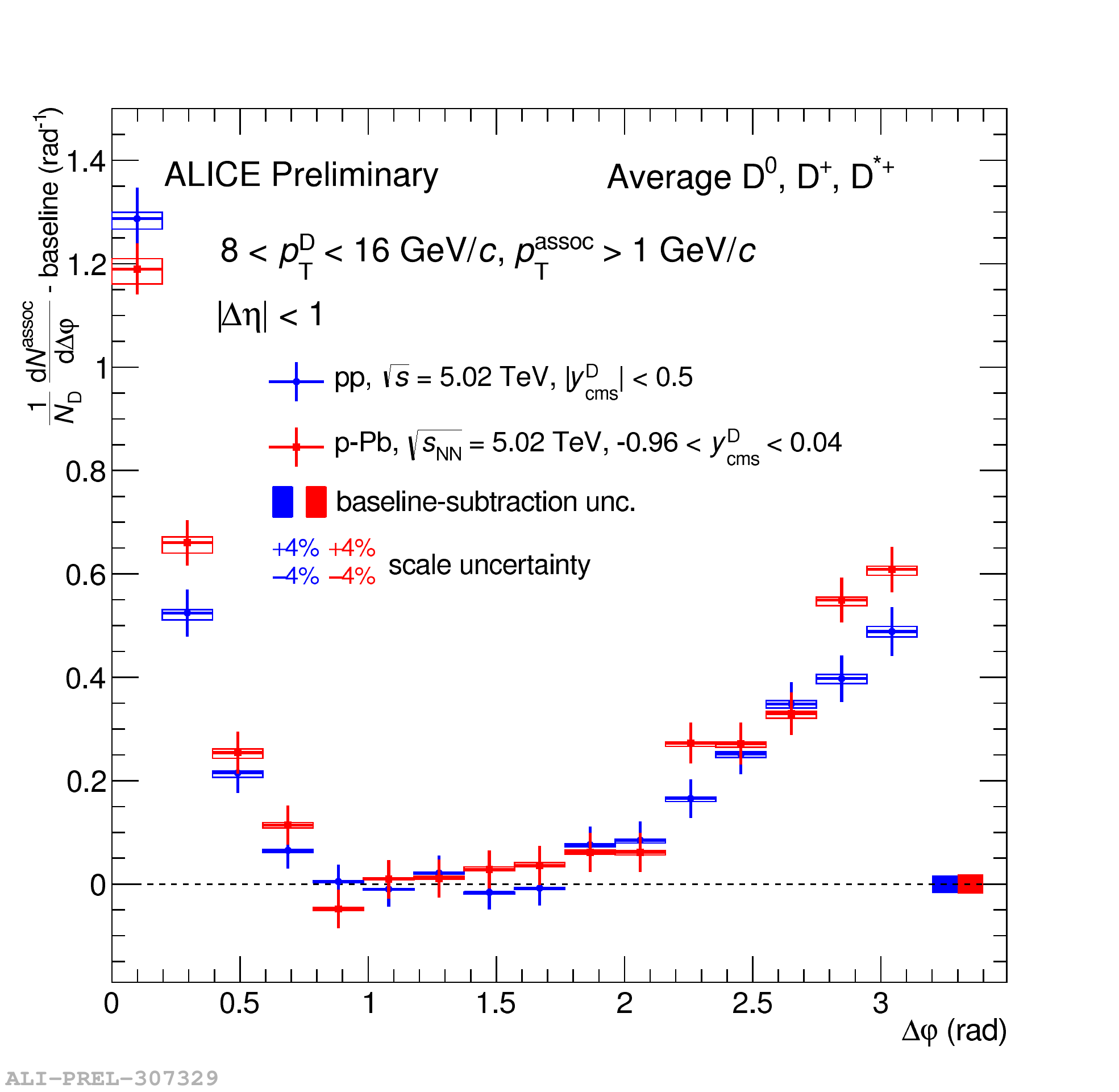}
	\end{center}
	\end{minipage}\hspace{1cm}
	\begin{minipage}{0.38\hsize} 
	\begin{center}
	\includegraphics[width=1\linewidth]{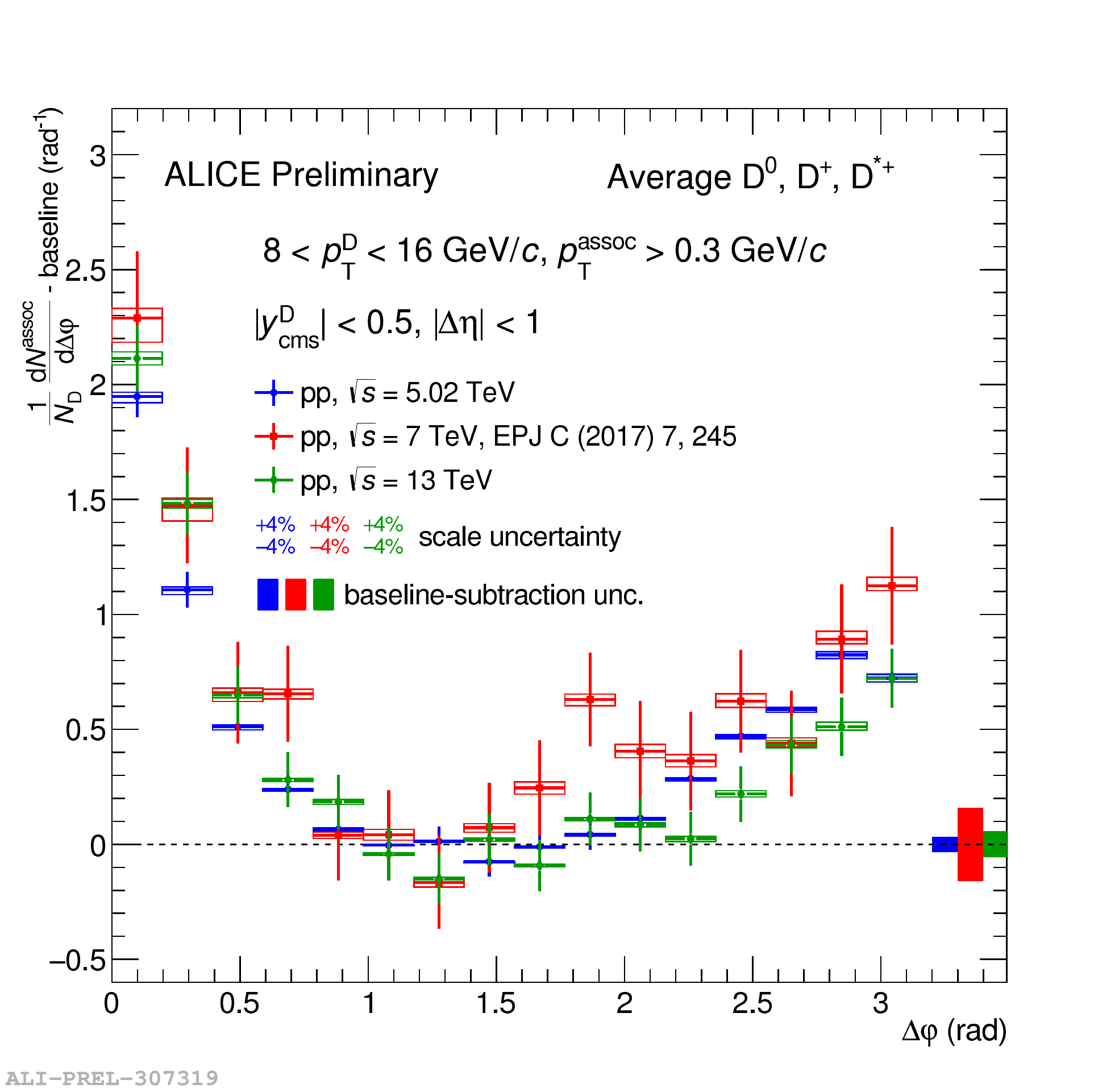}
	\end{center}
	\end{minipage}\vspace{-0.2cm}
	\caption{D-meson-hadron azimuthal correlations measured in pp and \pPb\ collisions at \sqrtsNN = 5.02 TeV (left), and in pp collisions at \sqrts = 5.02, 7, and 13 TeV (right).} 
	\label{fig:DmesonwithModel}
	\end{figure}

	Charm jet fragmentation is also studied using azimuthal ($\Delta \phi$) correlations between D mesons and associated charged hadrons. Two peaks are observed in the $\Delta \phi$ distribution, one at $\Delta \phi$ $\approx$ 0, called the near side, and a broader peak at the away side $\Delta \phi$ $\approx$ $\pi$, signifying two leading jets emitted in opposite directions in a collision. No evidence of CNM effects were found (see Fig.~\ref{fig:DmesonwithModel}, left panel) when production of associated tracks was compared across pp and p--Pb collisions, at $\sqrt{s_{\text{NN}}}$ = 5.02 TeV. No energy dependence could also be observed in pp collisions as seen on the right panel.
%%%%%%%%%%%%%%%%%%%%%%%%%%%%%%%%%%%%%%%%%%%%\vspace{-0.1cm}
\section{Summary}
ALICE obtained new measurements  of HFe jets in pp and p--Pb collisions at $\sqrt{s_{\text{NN}}}$ = 5.02 TeV. Also shown is a first measurement of D$^0$-jets in Pb--Pb collisions at $\sqrt{s_{\text{NN}}}$ = 5.02 TeV in 0--20\% centrality with a suppression by a factor of five at \pt\ $\sim$ 10 \GeVc. At $\sqrt{s}$ = 7 TeV, hard fragmentation for D$^0$-jets was seen in pp collisions in 5 $<$ \pt\ $<$ 15 GeV/$c$ along with a hint of softer fragmentation in 15 $<$ \pt\ $<$ 30 GeV/$c$ in data when compared to theoretical predictions. Our new measurements of b-tagged jets in p--Pb collisions at $\sqrt{s_{\text{NN}}}$ = 5.02 TeV were also reported. No energy dependence was observed for D-meson-hadron correlations in pp collisions. And there was no evidence of CNM effects in p--Pb collisions for any study reported in this manuscript.
\bibliographystyle{spmpsci_unsrt}
\bibliography{sqmbib.bib}
\end{document}